\documentclass[12pt]{article}

\input amssym.def
\input amssym
\def\be{\begin{equation}}
\def\ee{\end{equation}}

\begin{document}
\renewcommand{\theequation}{\arabic{section}.%
\arabic{equation}}

\title{Spacetimes, electromagnetism and fluids\\
(a revision of traditional concepts)\footnote{A
talk given at the IV Workshop
in Gravitation and Mathematical Physics,
November 25--30, 2001, Chapala, Jal., M\'exico.}}
\author{Nikolai V. Mitskievich\thanks{Departamento
de F{\'\i}sica, CUCEI, Universidad de Guadalajara,
Guadalajara, Jalisco, M\'exico;
e-mail: nmitskie@udgserv.cencar.udg.mx}}
\maketitle
\vspace*{-1.cm}
\begin{abstract}
In this talk $r$-form fields in spacetimes of any
dimension $D$ are considered ($r<D$). The
weak-field Newtonian-type limit of Einstein's
equations, in general, with relativistic sources
is studied in the static case yielding a revision
of the equivalence principle (intrinsically
relativistic sources generate twice stronger
gravitational fields and hyperrelativistic sources
--- {\em e.g.}, the stiff matter --- generate four
times stronger gravitational fields than
non-relativistic sources). It is shown that
analogues of electromagnetic field, strictly
speaking, exist only in even-dimensional
spacetimes. In (2+1)-dimensional spacetime, the
field traditionally interpreted as ``magnetic''
turns out to be in fact a perfect fluid, and
``electric'', a perverse fluid (this latter concept
arises inevitably in the $r$-form description of
fluids for any $D$, and we consider here perverse
fluids in (3+1)-dimensional spacetime too). New
exact solutions of (2+1)-dimensional Einstein's
equations with perfect and perverse fluids are
obtained, and it is shown that in this case there
exists a vast family of static solutions for
non-coherent dust, in a sharp contrast to the
(3+1)-dimensional case. New general interpretation
of the cosmological term in $D$-dimensional
Einstein's equations is given via the
($D-1$)-form field, and it is shown that this
field is as well responsible (as this is the
case in 3+1 dimensions) for rotation of perfect
fluids [($D-2$)-form fields], thus the ``source''
term in the corresponding field equations has to
be interpreted as the rotation term.
\end{abstract}

\section{Introduction} 

We know --- till now, mostly empirically --- that
our Universe is four-dimensio\-nal, with one
temporal and three spatial dimensions. Naturally,
there were some attempts to understand why this
is really the case, but these attempts usually
reduced to observation of how the ``real''
dimensionality affects the behaviour of particles
and fields via the properties of dynamical
equations, and testing strange peculiarities arising
from application of the same equations to cases with
other numbers of dimensions. However, in some
versions of unified field theories, as well as in
supersymmetry considerations, certain progress was
made in understanding how our standard (and
postulative) approach to the dimensionality of the
Universe could follow from higher-dimensional
theories (the compactification procedures also have
to be mentioned in this connection). Still, there is
no sign that any serious work was done in
trying to trace an evolutionary formation
of our 4-dimensional Universe from quantum
theoretical and cosmological considerations,
the formation which could begin with other number
of dimensions, probably, with $1+1$ (see some
remarks made in the last Section). In general,
on the probable ways of formation of the laws
of physics, a Chapter however was written by
Wheeler on his ``Pregeometry'' \cite{MisThorWhe},
though it did not produce any response during
so many years.

But, of course, there is not only dimensionality
that matters; the very signature of spacetime,
$+,-,-,-$ (in fact, already stated above when
$3+1$ and $1+1$ were written), until recently
remained to be an invariable and dumb postulate.
Of course, there exists (in the literature) the
twistor four-dimensional space, with its signature
$+,+,-,-$, so well fitting many principles of
special relativity and elementary particles
theory. The twistor theory practically remains to
be an isolated islet in the sea of the conventional
four-dimensional ($+,-,-,-$)\footnote{Or,
($-,+,+,+$), which is the same; see, however,
interesting comments in \cite{PenRin}.}
theory, even without any attempts to formally
apply to its geometry the mere classical scheme
of general relativity. Probably, the most important
breakthrough was achieved in studies of signature
by H.~van Dam and Y. Jack Ng \cite{vDamNg}. Few
months ago they have shown, using group-theoretical
ideas of Wigner, that the non-trivial spin spectrum
of particles in quantum theory is possible only
in $3+1$ dimensions (more generally, in $n+1$
dimensions). See also the references given in
\cite{vDamNg}.

In this talk I'll try to discuss some
non-evolutionary backgrounds which could be
of use in the realization of this ambitious
but, presumably, inevitable program. After all,
one has to clean the ground before starting to
dig the hole and lay a foundation.

Greek indices will be used for spacetime
$D$-dimensional coordinates while in the
$(D-1)$-dimensional spatial sections we use
Latin indices.

\subsection{{\boldmath $D=3+1$}: well established
 theory}

Almost everything we use today is based on the
four-dimensional classical physics, even the
quantum physics does not escape this destiny
(we do not know the intrinsic language of the
quantum world and have to apply the classical
basic concepts which naturally are subjected
there to the well known uncertainties). From
our four-dimensional experience we know that
one of the best studied and universally used
fields, the (Maxwell) electromagnetic one, is
(a) linear, (b) intrinsically relativistic
even when it is time-independent, (c) there is
a far-reaching analogy between its electric
and magnetic parts, which we describe via
the dual conjugation of the field tensor.
In this talk I will show that these three
properties are closely interrelated (see
Appendices \ref{A} and \ref{B}).

The Maxwell field is the vector field (its
potential is a 1-form). Thus it seems logical
to study other $r$-form fields. The scalar
(0-form) field was the first target for
physicists, and the Klein--Gordon equation was
derived already by Schr\"odinger as his first
step towards his famous non-relativistic wave
equation. The real mystery of the scalar field
is why it serves mainly as a simple and nice
example, but does not play any really
fundamental and central r\^ole in today's
physics (the scalar-tensor approach to gravity
and the non-linear scalar field in general are
clearly not directly relevant to the $r$-form
fields study).

Strangely enough, only quite recently the
2-form field theory was directly applied to
field theoretic description of perfect fluids
\cite{Mits99a, Mits99b}, interpreted as those
via the automatically realized specific form
of the stress-energy tensor,
\be  \label{TmunuFl}
T^{\rm pf}=(\mu+p)u\otimes u-pg.
\ee
The fact of so late understanding of the 2-form
fields in 3+1 can be in a certain (indirect)
sense related to the erroneous (but however
only recent) denial by Weinberg \cite{Weinb96}
of the physical significance of 2- and 3-form
fields (in particular, omitting the issue of
radically different dynamical properties of 0-
and 2-form fields, see \cite{Mits99b}). In this
talk I fill a substantial gap in \cite{Mits99a}
where an alternative (to the perfect fluid) case
in the 2-form field theory was not considered
--- the perverse fluid case, as I dare call it.
Its inclusion is of importance at least since
exactly this exotic case plays a significant
r\^ole in 2+1-spacetime (then, of course, in
the capacity of the 1-form field which describes
there both perfect and perverse fluids).

Finally, the 3-form field (I call it the Machian
field relating it also, somewhat arbitrarily, to
the fundamental hypothetical field proposed by
Sakurai \cite{Sakurai}) was introduced in
\cite{Mits99a, Mits99b}. In this talk the
($D-1$)-form field is considered which takes the
place of the 3-form field in 3+1, playing the
same r\^ole. This really exotic field, admitting
either constant or arbitrary functions as
solutions for its potential (thus having global
and not local properties in a contrast to all
other physical fields whose equations belong to
the hyperbolic type), plays two very distinct
r\^oles: (a)~as a free field, it is responsible
for existence or absence of the cosmological
constant (in the latter case, and only then,
the Machian field is intrinsically relativistic);
(b) an interaction between this field and the
2-form field (in 3+1) is the only means to impart
a rotation to the 2-form (fluid) field. Thus this
3-form field in 3+1-spacetime perfectly fits
in Ernst Mach's world picture.

\subsection{{\boldmath $D=2+1$}: hopes and
 prejudices}

The three-dimensional (2+1) spacetime was
repeatedly considered in many publications
primarily to the end of finding guidelines
of quantization of gravity, since the 2+1
case seemed to offer radical simplifications
both in the canonical description of gravity
and in the topological properties of the
model spacetime (see, {\em e.g.},
\cite{Carl95, GiAbKu}). Naturally, the
general attention was also attracted by
the classical (2+1)-geometry, exact solutions
of (2+1)-Einstein's equations, and in
particular by the (2+1)-black holes:
see \cite{CorFran} where Einstein's equations
were reformulated (I prefer to use their
traditional form below, without inclusion of
the dimensionality in the gravitational
constant); \cite{ChaChaMan, KamKoi, BaHeTeZa},
with an important correction in \cite{Garc}.
This last correction proved to be only the
beginning of a revision of the previous
general approach based on some na{\"\i}ve
prejudices concerning the electromagnetic
fields in $D$ dimensions, and the next (though
not the last) step of this revision was
\cite{MitGar}. 

Thus the vector field in 2+1
cannot be interpreted as electromagnetic field:
it describes instead the (perfect and perverse)
fluids. The case previously treated as electric
field, is in fact the 2+1 perverse fluid, while
the 2+1 ``magnetic field'' finds its physical
interpretation as the perfect fluid (in
\cite{BarBurLan} the authors came to a very
similar, though not sufficiently general
conclusion), exactly with the eigenvalues of
its stress-energy tensor as the fluid energy
density (corresponding to the eigenvector
which is timelike in this latter case) and
two equal pressures (isotropy) for any pair of
spacelike vectors on the spacetime section
orthogonal to the timelike eigenvector.
Moreover, the inhomogeneity (in the dynamical
field equations) whose presence was
previously interpreted as electric charge and
current distributions, now is proven to be
{\em not a source term}, but the rotation term
in the dynamical field equations; this
conclusion is supported by several logically
firm mathematical and physical arguments. This
rotation term is due to the interaction
between the 1-form and (Machian) 2-form fields
in 2+1 spacetime, in a complete analogy with
the situation in 3+1 where the ranks of the
respective $r$-form fields have to be
increased by one thus producing the room for
the usual Maxwell (1-form in 3+1) field.

\subsection{{\boldmath $D=n+1$}: the systematic
 approach}

In this talk a systematic and self-consistent
approach to electromagnetism and other
(including model) fields is generalized to
the $D=(n+1)$-dimen\-sional spacetimes. We shall
use the concepts and notations introduced in
Appendices \ref{A} and \ref{B}, in particular,
concerning intrinsically relativistic and
hyperrelativistic fields. The $r$-form fields
($r<D$), the respective field equations and
stress-energy tensors are considered in
Section \ref{Sec2} together with the
eigenvalues of these $T^\mu_\nu$'s (with more
details in other Sections dedicated to the
specific fields), which make it easier not
only to arrive at the physical interpretation
of these fields, but also to provide adequate
tetrad and vielbein bases indispensable in
finding the corresponding exact solutions of
Einstein's equations. The model fields (not
seeming to be as fundamental as Maxwell's
and Mach's fields) describing fluids are
studied in Section \ref{Sec3} (non-rotating
case) and further in Section \ref{Sec6}
(Subsection \ref{Rot}, including rotation);
not only the perfect fluids with all possible
equations of state are considered, but also a
new concept of the ``perverse fluid'' is
introduced. In Section \ref{Sec6} it is shown
that the free $(D-1)$-form fields are
equivalent to appearance of the cosmological
term in Einstein's equations, the case of
$\Lambda=0$ being the intrinsically relativistic
case of the $(D-1)$-form fields (seeming to be
as fundamental as the generalized Maxwell
fields). Then, in Section \ref{Sec5}, the
electromagnetic (Maxwell-type) fields are
generalized to all even-dimensional spacetimes
(in odd spacetime dimensions true
electromagnetic-type fields are absent).
Some new exact general relativistic solutions
are reviewed and obtained for the perfect and
perverse fluids in Section \ref{Sec7}; in
particular, a vast family of static non-coherent
dust solutions in 2+1 is found, and it is shown
that there are no rotating dust solutions in 2+1.
Finally, in Section \ref{Sec8}, the concluding
remarks are given.

\setcounter{equation}{0}
\section{Skew rank {\boldmath\lowercase{$r$}}
fields and their stress-energy tensors}\label{Sec2}

The potential of a free $r$-form field is
\be
A=\frac{1}{r!}A_{\alpha_1\ldots\alpha_r}
dx^{\alpha_1}\wedge\cdots\wedge dx^{\alpha_r},
\ee
a skew-symmetric tensor of rank $r$, and the
field tensor (intensity) is
\be
F=\frac{1}{(r+1)!}F_{\alpha_1\ldots\alpha_{r+1}}
dx^{\alpha_1}\wedge\cdots\wedge dx^{\alpha_{r+1}}=dA,
\ee
while the only invariant on which the
Lagrangian (density) depends, is
\be
I=F_{\alpha_1\ldots\alpha_{r+1}}F^{\alpha_1
\ldots\alpha_{r+1}}.
\ee

The free (but, in general, non-linear) $r$-form
field equations then are
\be  \label{rFEq}
\left(\sqrt{|g|}\frac{dL}{dI}F^{\alpha_1\dots
\alpha_r\mu}\right)_{,\mu}=0 ~ \Leftrightarrow
d\left(\frac{dL}{dI}\ast F\right)=0.
\ee

The stress-energy tensor following from
the Noether theorem (see \cite{Mits58,
Mits69, Mits99a}) takes the form
\be  \label{2.1}
T^\beta_\alpha=-L\delta^\beta_\alpha+2
\frac{\partial L}{\partial g^{\alpha\mu}}
g^{\beta\mu}=-L\delta^\beta_\alpha+2(r+1)
\frac{dL}{dI}F_{\alpha\mu_1...\mu_r}
F^{\beta\mu_1...\mu_r},
\ee
so that its trace reads
\be  \label{2.2}
T^\alpha_\alpha=-DL+2(r+1)I\frac{dL}{dI}.
\ee

Below, when this will be more convenient,
we shall use other letters to denote the
specific fields, their invariants, and the
corresponding Lagrangians. Since the concrete
determination of eigenvalues and eigenvectors
of the respective stress-energy tensors for
arbitrary $D$ and $r$ crucially depends on
these characteristics, we shall consider them
for some concrete values of $D$ and $r$ only,
and in the corresponding Sections. Here it
is however worth mentioning that the Machian
($D-1$)-form field always has only one
eigenvalue (equal to zero in the intrinsically
relativistic case), and its eigenvector is
completely arbitrary. The ($D-2$)-form field
modeling perfect fluids has one timelike
eigenvector corresponding to the single
eigenvalue and $D-1$ spacelike eigenvectors
with the same ($D-1$)-fold (degenerate)
eigenvalue; these eigenvectors are orthogonal
to the timelike one. The same type of field
describing perverse fluids, has the same
number of eigenvalues and linearly independent
($D$) eigenvectors, but the single eigenvalue
corresponds to a spacelike eigenvector, and
of the other $D-1$ eigenvectors orthogonal
to it, one is timelike. The Maxwell-type
field [existing only in even $D$, the
potential being ($D/2-1$)-form] possesses
two distinct eigenvalues, the both
($D/2$)-fold degenerate. Since the
stress-energy tensor is real and symmetric,
the system of its eigenvectors can be
orthonormalized to form a natural basis
in $D$ dimensions (though for null fields,
of course, this is not the case: there is
always a pair of real null eigenvectors with
mutual normalization, one of which determines
the null direction in which the field is
propagating: the specific property of the
spacetimes whose signature is $+,-,\dots,-$).

\setcounter{equation}{0}
\section{Fluids in field-theoretic
 description}\label{Sec3}

\subsection{Perfect fluids}

Like in 3+1, the stress-energy tensor (\ref{2.1})
of the ($D-2$)-form field in $D$ dimensions
reduces to 
\be
T^\beta_\alpha=2I\frac{dL}{dI}b^\beta_\alpha
-L\delta^\beta_\alpha
\ee
where
\be  \label{proj}
b^\beta_\alpha=\delta^\beta_\alpha-u_\alpha
u^\beta, ~ ~ b^\beta_\alpha u^\alpha=0=
b^\beta_\alpha u_\beta, ~ ~
u=\frac{f}{\sqrt{f\cdot f}}, ~ ~ f=\ast F;
\ee
$\ast$ being the Hodge star (dual conjugation),
while $f$ is for perfect fluids a timelike
covector, $f\cdot f>0$, thus $u\cdot u=+1$. Since
$b^\beta_\alpha$ is the projector on the (local)
subspace orthogonal to the congruence of $u$,
the latter is eigenvector of the stress-energy
tensor with the eigenvalue $(-L)$
\be
T^\beta_\alpha u^\alpha=-Lu^\beta,
\ee
while any vector orthogonal to $u$ is also
eigenvector, now with the ($D-1$)-fold eigenvalue
$2I\frac{dL}{dI}-L$. This is the property of
the stress-energy tensor of a perfect fluid
possessing the invariant mass density $\mu$
and pressure $p$ (in its local rest frame):
\be  \label{mu_p}
\mu=-L, ~ ~ p=L-2I\frac{dL}{dI}
\ee
[the eigenvalue corresponding to the spacelike
eigenvectors, is ($-p$)].
						 
The free ($D-2$)-form field equations are
\be  \label{2formFE}
d\left(I^{1/2}\frac{dL}{dI}u\right)=0.
\ee
Thus the free $r=D-2$ field case can describe
only non-rotating fluids.

Perfect fluids characterized by the simplest
equation of state
\be   \label{eqstate}
p=(2k-1)\mu,
\ee 
correspond to the Lagrangian $L=-\sigma|I|^k$,
$\sigma>0$. In 3+1, the important special cases
are: the incoherent dust ($p=0$) for $k=1/2$,
intrinsically relativistic incoherent radiation
($p=\mu/3$) for $k=2/3$, and hyperrelativistic 
stiff matter ($p=\mu$) for $k=1$. There are two
ways to consider the property of the fluid to be
intrinsically relativistic and hyperrelativistic:
(a) from the point of view of the relation
between the temporal and spatial parts of the
stress-energy tensor (essentially, the sign of
its trace), this approach to be used below in
this talk (see Table 1 in the
Appendix \ref{B}), or (b) taking into account
the coefficients in the terms proportional to
$\mu$ and $p$ in (\ref{2.1.8a}), though in
this case the (2+1)-spacetime obviously falls
out of the consideration. In the approach (a)
it is remarkable that the electromagnetic
fields in even dimensions where they only
exist, are intrinsically relativistic when
their equations are linear, like this occurs
in 3+1.

One may similarly treat polytropes
($p=A\mu^\gamma$), though in this case the
Lagrangian is determined only implicitly,
like this is known for $D=3+1$, \cite{Mits99a}.

In \cite{Mits99a} it was found that in the
special relativistic approximation (the only
approximate case considered in \cite{Mits99a}),
the low-amplitude mass density perturbations
of a relativistic perfect fluid (thus the
relativistic sound) propagate with exactly the
same velocity as it is known in
the non-relativistic phenomenological theory.
Using the traditional polytrope state equation,
one gets then for the acoustic velocity, for
example, in the air the standard expression
$c_s=\sqrt{\gamma p/\mu}$. The same approach
applied to the equation of state (\ref{eqstate})
permits to naturally introduce the concept of
the stiff matter equalizing the acoustic
velocity to that of light (in our units, $c=1$).
These conclusions are universal for all $D$'s.

\subsection{Perverse fluids}

This is the case when $f\cdot f<0$ (the
(co)vector dual to the field tensor is
spacelike), so that it can be normalized as
\be
l=\frac{f}{\sqrt{-f\cdot f}}, ~ ~ l\cdot l=-1.
\ee
It was just mentioned in \cite{Mits99a} as the
tachyonic (abnormal) fluid. The stress-energy
tensor of this ($D-2$)-form field reads
\be   \label{Ttach}
T^\beta_\alpha=-L\delta^\beta_\alpha+
2I\frac{dL}{dI}\tilde{b}^\beta_\alpha, ~ ~
\tilde{b}^\beta_\alpha=\delta^\beta_\alpha+
l^\beta l_\alpha
\ee
where $\tilde{b}^\beta_\alpha$ is the projector
on the local subspace orthogonal to the $l$
congruence. It is clear that this subspace is
timelike, and every vector in it is eigenvector
of $T^\beta_\alpha$ with one and the same
eigenvalue ($-\tilde{p}$), while $l$ is an
eigenvector corresponding to the non-degenerate
eigenvalue $\tilde{\mu}$. These eigenvalues
have the same structure as those in
(\ref{mu_p}) which is then rewritten with
tildes, but they now pertain to another
combination of eigenvectors, so that a {\em
tilde} is put over the letters denoting them.
Below an example of the (3+1)-spacetime is
considered.

If we orthonormalize the three linearly
independent eigenvectors related to $\tilde{p}$,
and admit $l$ as the fourth unit vector, the
natural tetrad is formed with respect to which
the description of a perverse fluid should look
most simple. There is also a symmetry (isotropy)
in the local section orthogonal to $l$ which
may be of use in simplifying the field equations;
this suggests, in particular, that a kind of
rotation could be most probably introduced
(if one looks for rotating solutions) which
involves a combination of $l+\omega dt$, and
not $dt+\omega d\phi$ as this is the case for
rotating perfect fluids. Let us write the
tetrad $\theta^{(\alpha)}$ so that the first
($\alpha=0$) covector, as well as the next two,
will correspond to the local subspace orthogonal
to $l$, while $\theta^{(3)}=l$. Then the
stress-energy tensor will take the diagonal form
\be
T_{(\alpha)(\beta)}\theta^{(\alpha)}\otimes
\theta^{(\beta)}=\label{Ttprop}
\tilde{p}\left(\theta^{(0)}
\otimes\theta^{(0)}+\theta^{(1)}\otimes
\theta^{(1)}+\theta^{(2)}\otimes\theta^{(2)}
\right)+\tilde{\mu}l\otimes l.
\ee

The issue of the equation of state of perverse
fluids is very much the same as that of perfect
fluids (see above and in \cite{Mits99a}). Some
differences however arise in the consideration
of propagation of perturbations (special
relativistic theory). It can be considered
with 0th approximation for $l$ being either
$dz$ or $rd\phi$ (in the last case, a transition
to the Cartesian coordinates can be carried out
globally, but, since this approach is an
approximation to general relativistic theory,
at any point being not at the origin, thus
$r\neq0$, one can take other tangent frame in
a vicinity of that point, and pretend to use
there Cartesian frame; the only exception is
the ``singular'' point at the origin which we
shall not discuss here). So let us consider
$f=dz+\delta f$ and $I=-1-2\delta f_z$ (the
second-order term is neglected here and
subsequently). The pseudopotential $\Phi$ can
be introduced due to the field equation
(\ref{2formFE}) where, of course, $I$ should
be changed by ($-I$) for a perverse fluid.
Thus
\[
d\Phi\equiv\frac{dL}{d(-I)}f=
\left[\frac{dL}{d(-I)}dz+\frac{dL}{d(-I)}
\delta f+2\frac{d^2L}{d(-I)^2}
\delta f_zdz\right]_{I=-1},
\]
the expression which yields two equations,
\[
\left[\frac{dL}{d(-I)}+2\frac{d^2L}{d(-I)^2}
\right]_{I=-1}(\delta f_z)_{,a}=\left[
\frac{dL}{d(-I)}\right]_{I=-1}
(\delta f_a)_{,z}
\]
and
\[
\left[\frac{dL}{d(-I)}\right]_{I=-1}(\delta
f_a)_{,b}=\left[\frac{dL}{d(-I)}\right]_{I=-1}
(\delta f_b)_{,a}
\]
($a,b=0,1,2$) where the indices $a$ and $b$
pertain to the timelike section, thus not
containing $z$-component. The last equation
is satisfied when
\[
\delta f_a=\left[\frac{dL/dI+2d^2L/dI^2}{d
L/dI}\right]_{I=-1}\left(\int\delta f_zdz+
\phi\right)_{,a};
\]
there are two functions, $\delta f_z$ and
$\phi(t,x,y)$, which are still undetermined.
We use now the fact that $\delta f$ (as well
as $f$ itself) is divergenceless: this means
that
\[
-\delta f^z_{,z}=-\delta f^a_{,a}=
-\delta f_{t,t}+\delta f_{x,x}+
\delta f_{y,y}=
\]
\[
\left[\frac{dL/dI+2d^2L/dI^2}{d
L/dI}\right]_{I=-1}\tilde{\Delta}\left(\int
\delta f_zdz+\phi(t,x,y)\right)
\]
where $\tilde{\Delta}=\partial^2/\partial x^2+
\partial^2/\partial y^2-\partial^2/\partial t^2$
is an analogue of the Laplacian [or,
truncated D'Alembertian] operator (in a timelike
hypersurface of subspace with coordinates $x^a$).
Taking the derivative of the both sides of the
last relation with respect to $z$, we arrive at
\[
\frac{\partial^2\delta f_z}{\partial z^2}+
\left[\frac{dL/dI+2d^2L/dI^2}{d
L/dI}\right]_{I=-1}\tilde{\Delta}\delta f_z=0.
\]
But it is worth dividing this equation by the
constant coefficient before $\tilde{\Delta}$
in order to directly see with what velocity
do propagate the perturbations in different
directions in this obviously anisotropic world;
then we get the squared velocity as a
coefficient before the second partial derivative
with respect of the corresponding spatial
coordinate (the coefficient before $\tilde{\Delta}$
will be equal to unity).
Thus we finally find that
\be
\left\{\left[\frac{dL/dI}{dL/dI+2d^2L/dI^2}
\right]_{I=-1}\frac{\partial^2}{\partial z^2}+
\tilde{\Delta}\right\}\delta f_z=0.
\ee
Consequently, the perturbations propagate in
a perverse fluid in all directions except that
which corresponds to the non-degenerate eigenvalue,
with the velocity of light, while in this last
(here, $z$) direction this velocity is inverse
to the acoustic one being characteristic to a
perfect fluid with the same equation of state,
now --- for the perverse fluid --- applied to
$\tilde{p}$ and $\tilde{\mu}$ (the velocities are
given in the units of the velocity of light,
thus they are dimensionless). There is still
another distinction from the perfect fluid case:
for the perverse fluid the perturbation whose
propagation is considered, is {\em not} that of
the mass density, but of the $z$ component of the
anisotropic pressure (here, $\tilde{\mu}$).

We see that the only concept which remains
unchanged is that of the stiff matter (with the
velocity of light, $c=1$, for the acoustic-type
waves in this fluid). Otherwise, in order to
obtain subluminal velocities, one has to consider
other part of the range of the parameter $k$
which for a perfect fluid would correspond to
crucially unphysical cases outside the stiff
matter states. This fact is in a complete
agreement with the interpretation of perverse
fluids as ``tachyonic'' fluids. The ``only''
hard question in this interpretation is why
a particular concrete spatial direction (here,
$z$) is singled out, when if a stochastic motion
of tachyons is being considered, there should be
a spatial isotropy in the overall picture. The
theory, especially if we go to the mentioned
``other part'' of the range of the parameter
$k$, is highly nonlinear, and there is no way
to take a meaningful superposition of all
possible directions of the axes here named as $z$.
This is clearly a conflict between the usual
concepts of physical objects and the tachyonic
ones; the only case which can be meaningfully
considered in this connection, is the case of
linear equations, with $k=1$, when a superposition
of solutions is automatic. Probably, from this
starting point one should begin the formulation
of the statistical model of perverse fluids.
Otherwise these theoretical objects should be
taken as some formal concepts, though sufficiently
well described in general relativistic field
theory, and one has to look for further results
following from this theory in order to come to
a better understanding of this subject.

An amazing situation has been however developed
in the (2+1)-spacetime theory where one of the
best studied solutions is that with a perverse
fluid (though generally misinterpreted as an
``electric field'').

\subsection{Null fluids (coherent radiation)}

In the null, or radiation case the vector $f$
is null, $I=0$, and the respective eigenvalue is
equal to zero. The stress-energy tensor cannot
be brought to a diagonal form, it always contains
a nontrivial flux component. However this component
can be made as small as one wishes (the Doppler
effect), only its vanishing occurs for the
degenerate (forbidden) transformation to the
velocity of light in the direction of the flux.

In fact, the stress-energy tensor of the null
fluid reads $T^\beta_\alpha=
\lambda f_\alpha f^\beta$. Now $f$ is not a simple
eigenvector: there exist other $D-2$ eigenvectors
corresponding to the eigenvalue zero, and these
are spacelike vectors (say, $l_a$, $a=3,\dots,D$)
orthogonal to $f$. Let us rename $f$ as $v$; thus
$v\cdot v=0$, $l_a\cdot l_b=-\delta^a_b$,
$v\cdot l_a=0$, while
\be
T^\beta_\alpha=\lambda v_\alpha v^\beta, ~ ~
T^\beta_\alpha v^\alpha=0=T^\beta_\alpha l^\alpha_a.
\ee
We choose the last vector, $w$, to be null, orthogonal
to $l_a$, and normalized with respect to $v$ as
$w\cdot v=1$. Then the two mutually related bases
(one covector, $\theta^{(\alpha)}$, and another vector
one, $X_{(\alpha)}$) can be introduced as
\be
\theta^{(0)}=\underline{w}, ~ \theta^{(1)}=
\underline{v}, ~ \theta^{(a)}=\underline{l}_a
\ee
and
\be
X_{(0)}=\overline{v}, ~ X_{(1)}=\overline{w}, ~
X_{(2)}=-\overline{l}_a,
\ee
thus $\theta^{(\beta)}\cdot X_{(\alpha)}=
\delta^\beta_\alpha$ (the underlined objects
being covectors and overlined, vectors).

In anticipation, it may be mentioned that here the
rotating field can easily be described, as this is
the case for the perfect and perverse fluids. Then
the (co)vector $w$ will acquire an additional term
proportional to $v$ (the coefficient being a function
of coordinates), {\em cf.} \ref{Rot}.

\setcounter{equation}{0}
\section{The Machian field}\label{Sec6}

The $(D-1$)-form field will be described in $D$
by the potential
\be
C=\frac{1}{(D-1)!}C_{\mu_1\dots\mu_{D-1}}
dx^{\mu_1}\wedge\dots\wedge dx^{\mu_{D-1}}
\ee
yielding the skew-symmetric field tensor
($D$-form)
\be
W=dC.
\ee
Its invariant reads
\be     \label{K}
K=\ast(W\wedge\ast W)\equiv\frac{1}{D!}
W_{\alpha_1\dots\alpha_D}W^{\alpha_1\dots\alpha_D}
\equiv\tilde{W}^2;
\ee
we shall use it to construct the Lagrangian of
a free ($D-1$)-form field. Here
\be \label{tildeW}
\tilde{W}:=\ast W=\frac{1}{D!}W_{\alpha_1\dots
\alpha_D}E^{\alpha_1\dots\alpha_D}, ~ \mbox{thus}
~ ~ W_{\alpha_1\dots\alpha_D}=:\tilde{W}
E_{\alpha_1\dots\alpha_D},
\ee
$\tilde{W}$ being axial scalar (often called
``pseudoscalar'').

\subsection{The cosmological term} \label{Lambda}

For the ($D-1$)-form field with its invariant
(\ref{K}), the stress-energy tensor takes the form
\be
T^\beta_\alpha=\left(2K\frac{dL}{dK}-L\right)
\delta^\beta_\alpha,    \label{T2ff}
\ee
exactly coinciding with this tensor for the 3-form
field in 3+1 (but now with $D$-dimensional
indices). A similar situation repeats for the
field equations: they reduce to
\be
K^{1/2}\frac{dL}{dK}=\mbox{const}.  \label{Eq2ff}
\ee
The further conclusions are the same as in
\cite{Mits99a} for the 3-form field, and we repeat
them here in short. If $L\sim K^{1/2}$, all
($D-1$)-forms $C$ identically satisfy (\ref{Eq2ff}),
while the stress-energy tensor identically vanishes.
Otherwise, $K$ should be constant, and the
stress-energy tensor (\ref{T2ff}) becomes
proportional to $\delta^\beta_\alpha$ with a
constant coefficient obviously identifiable with
the cosmological constant $\Lambda$. The case
$L\sim K^{1/2}$ corresponds to $\Lambda=0$, this
being the intrinsically relativistic ($D-1$)-form
field in $D$, offering an alternative interpretation
to the cosmological constant problem (see
\cite{Mits99a}). Since the 2-form $C$ also permits
to introduce rotating fluids in 2+1 (similarly
to this role of 3-form field in 3+1, though we leave
here this subject without further consideration), we
are inclined to relate this field to the fundamental
cosmological Machian field (probably, of the type
of that proposed by Sakurai, \cite{Sakurai}).

\subsection{Rotating systems}\label{Rot}

Turning to the rotating fluid case, one has to
generalize the dynamical equation, for example
that previously taken in the form (\ref{2formFE}),
so that it will describe a rotating congruence
$f$ or, equivalently, $u$. The (3+1)-dimensional
case was already discussed in
\cite{Mits99a, Mits99b}. Below we shall consider
only the case of the (2+1)-spacetime where
we have to add to the Lagrangian $L(I)$ a
new term, say, $M(J)$, where a new invariant
involving $\tilde{W}$ [see (\ref{tildeW})]
as well as $A_\mu f^\mu$, reads
\be  \label{J}
J:=W_{\lambda\mu\nu}A^\lambda F^{\mu\nu}\equiv
\tilde{W}E_{\lambda\mu\nu}A^\lambda F^{\mu\nu}
\equiv2\tilde{W}A^\lambda f_\lambda
\ee
[a product of two axial scalars (pseudoscalars)].
Due to the complete antisymmetrization of the
product $A\wedge F$ in (\ref{J}) involving three
indices, the additional term appearing in the
stress-energy tensor, will be of the type of
(\ref{T2ff}):
\be
T^\beta_\alpha=\left(2J\frac{dM}{dJ}-M\right)
\delta^\beta_\alpha,
\ee
thus it will vanish if $M\sim J^{1/2}$, or if
$J\frac{dM}{dJ}$ vanishes simultaneously
with $M$ due to some property of $M$, as
this will be the case for the problems considered
in Section \ref{Sec7}. The new 1-form field
equation is
\be  \label{new1ff}
d\left(\frac{dL}{dI}f+\frac{dM}{dJ}\tilde{W}A
\right)=-\frac{dM}{dJ}\tilde{W}F,
\ee
thus $d\left(\frac{dM}{dJ}\tilde{W}F\right)=0$
is a condition on $\tilde{W}$.

For the 2-form field, the additional term in the
left-hand side of the equation (\ref{Eq2ff}), emerges:
\be
2\frac{dM}{dJ}A^\lambda f_\lambda
  \label{addeq2ff}
\ee
[its sum with the left-hand side of (\ref{Eq2ff})
to be a constant; however, this contribution vanishes
at least under a sufficiently general class of
natural assumptions, so we do not write here the
complete sum of terms from (\ref{Eq2ff}) and
(\ref{addeq2ff})]. Thus from (\ref{new1ff}) and
(\ref{Eq2ff}) plus (\ref{addeq2ff}) one can calculate
the function $\tilde{W}$ which was in fact still
arbitrary, and obtain the rotation characteristics of
the fluid.  This completes the introduction of
rotation in the theory of perfect fluids in 2+1
({\em cf.} a different method used in 3+1,
\cite{Mits99a}; the important differences are
here due to the change of dimensionality).

\setcounter{equation}{0}
\section{Electromagnetism ~ from ~ the
 systematic viewpoint}\label{Sec5}

As in the case of a perfect fluid, we postulate
here in $D$ dimensions essentially the same
properties of the stress-energy tensor as they
are in 3+1 for the field under consideration
(now, Maxwell's field). The only differences
are those which are due to the other dimensionality.
It is easy to see that the same distribution of
eigenvalues and eigenvectors (the degenerate pairs
of eigenvalues, and $D$ eigenvectors equally
divided between these eigenvalues) is possible only
in even number of dimensions, thus $D=2m$. This
means that the intensity tensor $F$ of the field
should be an $m$-form [hence the potential tensor
$A$ is an ($m-1$)-form], and in accordance with the
terminology of the theory we call this field the
($m-1$)-form field. In general, this field gives
two invariants, $I_1=F\cdot F$ and $I_2=F\cdot\ast
F$. $\ast$ means here the Hodge star (dual
conjugation) which merely rearranges the components
of $F$ when passing to $\ast F$, as this is the case
in 3+1. Similarly, the second invariant in general
reduces to a covariant divergence, thus not being
of use in obtaining the linear Euler--Lagrange
equations (it can result then merely in a surface
term). The analogy with the usual Maxwell theory
further spreads to the dynamical and structure field
equations. The stress-energy tensor is equally
distributed between two $m$-dimensional subspaces,
and in the linear theory ($k=1$) it manifests
intrinsically relativistic features --- always
when this Maxwell-type field can exist (for all
$D=2m$, see Table 1). It is obvious
that these properties cannot occur by a coincidence,
so this should express a profound law of nature.
One has all reasons to admit that this is a
fundamental field which singles out the
even-dimensional spacetimes in the physical picture
of universe.

In 1+1, and only then, the Maxwell-type field is
simultaneously a fluid field (its magnetic type
coinciding with a perfect, and electric, with a
perverse fluid, the both being intrinsically
relativistic), and, moreover, this is a scalar
field. In 2+1 and 4+1 (as in all other
odd-dimensional spacetimes) the Maxwell-type field
simply does not exist (see the Theorem in the
Appendix \ref{B}).

\setcounter{equation}{0}
\section{Some exact solutions}\label{Sec7}

In our approach some new ways can be used to
find exact Einstein--Maxwell and Einstein--Euler
solutions. In particular, when fluids are
described via ($D-2$)-form fields, the equations
of state corresponding to the specific
invariant-dependence of Lagrangians [$L(I)$]
provide additional algebraical equations which
give new relations between unknown functions.
Another way leading from already known solutions,
for example, in 3+1, to lower-dimensional (say,
2+1) solutions, is the use of spacetime sections
of the former spacetimes, sometimes with
a redetermination of functions (see \cite{MitGar}
as well as the null solutions given below).
Still another method of finding new exact
solutions, this time in higher-dimensional
spacetimes, is a hybridization (sometimes,
inbreeding- and chimaera-engineering) of
already-known lower-dimensional solutions or
their sections, in general involving a
redetermination of functions.

Below some examples are given of new solutions in
2+1, the incoherent dust and coherent null fluid
solutions.

\subsection{Generalities}

First we consider the general Einstein-Euler
equations with a natural metric ansatz.
The case when $I>0$, as we already know,
describes perfect fluids. In the curvature coordinates
(Synge's terminology) the appropriate orthonormal
tetrad basis is
\be   \label{prffltetr}
\theta^{(0)}=e^\alpha(dt-\Phi d\phi), ~ \theta^{(1)}=
e^\beta dr, ~ \theta^{(2)}=rd\phi,
\ee
$\alpha$, $\beta$, and $\Phi$ being functions of
$r$. Then
\be   \label{pfmetric}    
ds^2=e^{2\alpha}(dt-\Phi d\phi)^2-e^{2\beta}dr^2
-r^2d\phi^2, ~ ~ \sqrt{g}=re^{\alpha+\beta}.
\ee
It is easy to calculate (using, for example, the
Cartan exterior forms formalism) the Ricci tensor
components,
$$
R_{(0)(0)}=
-\frac{1}{r}\left(r\alpha'e^{\alpha-\beta}\right)'
e^{-(\alpha+\beta)}-\frac{1}{2r^2}\Phi'^2
e^{2(\alpha-\beta)},
$$
$$R_{(1)(1)}=
\left(\alpha'e^{\alpha-\beta}\right)'e^{-(\alpha+
\beta)}-\frac{\beta'}{r}e^{-2\beta}-
\frac{1}{2r^2}\Phi'^2e^{2(\alpha-\beta)},
$$
$$R_{(2)(2)}=
\frac{1}{r}\left(e^{\alpha-\beta}\right)'
e^{-(\alpha+\beta)}-\frac{1}{2r^2}\Phi'^2
e^{2(\alpha-\beta)},
$$
$$R_{(0)(2)}=
\frac{1}{2}\left(\frac{1}{r}\Phi'e^{(3\alpha-\beta)}
\right)'e^{-(2\alpha+\beta)},
$$
as well as the scalar curvature $R=R_{(0)(0)}-
R_{(1)(1)}-R_{(2)(2)}$. Of course, the curvature
coordinates are not applicable when Nariai-type
spacetimes are considered. The convenience of
the proper (eigenvector) basis is that with respect
to it the stress-energy tensor takes the diagonal
form, hence $R_{(0)(2)}=0$ and $R_{(1)(1)}=
R_{(2)(2)}$. The both of these
equations are easily integrated yielding
\be   \label{Phi}
\frac{1}{r}\Phi'e^{3\alpha-\beta}=\omega=
\mbox{const}
\ee
and
\be   \label{11-22}
\frac{1}{r}\left(e^\alpha\right)'e^{-\beta}=C=
\mbox{const},
\ee
respectively.

The remaining Einstein equations read
\be    \label{G00}
G_{(0)(0)}\equiv\frac{1}{2}\left(R_{(0)(0)}+
R_{(1)(1)}+R_{(2)(2)}\right)=-\varkappa\mu
\ee
and
\be    \label{Gii}
\frac{1}{2}\left(G_{(1)(1)}+G_{(2)(2)}\right)\equiv
\frac{1}{2}R_{(0)(0)}=-\varkappa p.
\ee
In the description of a perfect fluid they are
usually treated as ``definitions'' of the mass
density and pressure (though when a specific
equation of state is assumed, their combination
gives an additional restriction on the metric
coefficients).

Let us consider for a while the flat 2+1
spacetime in polar coordinates,
\be
ds_0^2=dt^2-dr^2-r^2d\phi^2,
\ee
with
\be
A=qa(r)d\phi
\ee
where $q$ is a constant, and see which
function $a(r)$ would fulfil the 1-form field
equation in a non-self-consistent problem
(without taking into account Einstein's
equations). First of all we observe that
\be
F=dA=qa'dr\wedge d\phi,
\ee
so that
\be
I=\frac{1}{2}F_{\mu\nu}F^{\mu\nu}=
\frac{q^2a'^2}{r^2}>0
\ee
(clearly, the perfect fluid case). Let also
$L=\sigma I^k$, thus
\be
\sqrt{g}\frac{dL}{dI}F^{\mu\nu}=\sigma k
\left(\frac{qa'}{r}\right)^{2k-1}\left(\delta^\mu_r
\delta^\nu_\phi-\delta^\mu_\phi\delta^\nu_r
\right).
\ee
Then the 1-form field equation reads
\be    \label{eqf'}
\left(\sqrt{g}\frac{dL}{dI}F^{\mu\nu}\right)_{,\nu}
=-\sigma k\left[\left(\frac{qa'}{r}\right)^{2k-1}
\right]'\delta^\mu_\phi=0
\ee
(no rotation is involved). The solution is $a'=r$
(the integration constant is incorporated into $q$);
another possible solution is $k=1/2$, but this is
the case of an incoherent dust which will be later
discussed separately, so we shall now consider
the first alternative.

Now, returning to our more general problem, we
plausibly postulate the 1-form field potential and
the corresponding field tensor to be
\be   \label{ftensor}
A=\frac{1}{2}qr^2d\phi, ~ ~ F_{\mu\nu}=qr\left(
\delta^r_\mu\delta^\phi_\nu-\delta^\phi_\mu
\delta^r_\nu\right)
\ee
(the second integration constant in $a(r)$ --- see
(\ref{eqf'}) --- corresponds to addition of an exact
form to $A$ and is dropped). This ansatz is similar to
the Horsk\'y--Mitskievich method of constructing
exact charged solutions in 3+1 (see \cite{HorMits}
and \cite{Steph}), though the Killing vector $\xi$
approach is not automatically applicable directly to
the 1-form field potential $A$ due to the (in general)
nonlinear nature of the field.

In (\ref{ftensor}) we find components of the covariant
field tensor in the spacetime metricized by
(\ref{pfmetric}); its independent nontrivial contravariant
components are
\be
F^{12}=\frac{q}{r}e^{-2\beta}, ~ ~
F^{10}=\frac{q}{r}\Phi e^{-2\beta}.
\ee
Since $f=\ast F$,
\be     \label{fmu}
f_\mu=qe^{\alpha-\beta}\left(\delta^t_\mu-\Phi
\delta^\phi_\mu\right), ~ ~ f^\mu=qe^{-(\alpha+\beta)}
\delta^\mu_0.
\ee
This means, in particular, that
\be
A_\mu f^\mu=0   \label{Aforthog}
\ee
[{\em cf.} the remarks on vanishing of $M$ in the
Subsection \ref{Rot}, as well as the expression
(\ref{addeq2ff})]. The relativistic velocity of the
fluid is the normalized vector $f$,
\be
u=\theta^{(0)}=e^\alpha(dt-\Phi d\phi), ~
~ u^\mu=e^{-\alpha}\delta^\mu_0,
\ee
and the field invariant is
\be    \label{7.10}
I=\frac{1}{2}F_{\mu\nu}F^{\mu\nu}=f_\mu f^\mu=
f_0f^0=q^2e^{-2\beta}.
\ee
This expression as well as those to be deduced
below will prove to be of importance in finding
exact solutions of the complete system of field
equations. The usual treatment of perfect fluids
concentrates essentially on Einstein's equations,
but we shall see that inclusion of 1- and 2-form
field equations and concrete expressions of
Lagrangian and stress-energy tensor components,
makes calculations much easier.

\subsection{Rotating solutions}

As to the Lagrangians and field equations, we
admit the relations given in Subsection \ref{Rot}
with vanishing invariant $J$ whose derivatives
with respect to $A_\mu$ and $F_{\mu\nu}$ are
however different from zero. Thus the 2-form
field equation [(\ref{Eq2ff}) plus (\ref{addeq2ff})]
is satisfied trivially (we shall not use the free field
Lagrangian in this case) and the 1-form field
equation takes the form
\be
d\left[\frac{dL}{dI}2e^{\alpha-\beta}(dt
-\Phi d\phi)+\frac{dM}{dJ}\tilde{W}r^2d\phi\right]=
-2\frac{dM}{dJ}\tilde{W}r\,dr\wedge d\phi.
\ee
This equation implies (for general $a(r)$ as well)
\be     \label{Euler1}
\frac{dL}{dI}e^{\alpha-\beta}=-N=\mbox{ const}.,
\ee
a new integral when $\Phi\neq 0$,
and
\be     \label{Euler2}
\left(\frac{dM}{dJ}\tilde{W}r^2+2N\Phi\right)'=
-2\frac{dM}{dJ}\tilde{W}r.
\ee
It is clear that rotation will not disappear only if
$M=\lambda J$ with $\lambda=$ const; we put
$\lambda=1$ without infringing generality of
our considerations. Thus
\be     \label{8.14}
\tilde{W}=-\frac{2N}{r^4}\int\Phi'r^2dr,
\ee
while the function $\Phi$, as well as another
unknown function, have to be found from
Einstein's equations. In 3+1 we call the conditions
corresponding to (\ref{Euler1}) and (\ref{Euler2})
in the Horsk\'y--Mitskievich approach, the Maxwell
conditions (they follow there from Maxwell's
equations); let us baptize them in 2+1 as the Euler
conditions.

Turning to the simplest equation of state
(\ref{eqstate}) when $L=-\sigma I^k$, thus
$dL/dI=kL/I$, and taking into account (\ref{7.10})
and (\ref{Euler1}), we find that
\be    \label{Lmu}
\mu=-L=\frac{Nq^2}{k}e^{-\alpha-\beta}.
\ee
Thus
\be
p=\frac{2k-1}{k}q^2Ne^{-\alpha-\beta}.
\ee
However there is another way to express $L$:
directly from (\ref{7.10}),
\be   \label{Lgamma}
L=-\sigma I^k=-\sigma q^{2k}e^{-2k\beta}.
\ee
Comparing the two expressions of $L$, (\ref{Lmu})
and (\ref{Lgamma}), we come to a remarkable
algebraic relation between $\alpha$ and $\beta$,
\be    \label{alphbet}
\exp(\alpha-(2k-1)\beta)=\frac{2N}{2k\sigma}
q^{2(1-k)}.
\ee
This means that we need to determine only two
functions to solve our problem, say, $\alpha$ and
$\Phi$ (in the non-rotating case there remains
only one function to be found).

\subsection{Incoherent dust}

Incoherent dust ($k=1/2$, thus $p=0$) is here a
special case for which it follows from (\ref{alphbet})
that $\alpha=$ const, but this, according to (\ref{Gii}),
excludes the possibility of rotation for the admitted
form of the metric, with the both $\omega$ and $C$
being equal to zero. Thus in a 2+1 spacetime there
exists a continuous family of {\em static} non-rotating
dust solutions (among them a continuous set of
singularities-free ones), in fact with an arbitrary
spherically symmetric distribution of mass density, in
an acute contrast to the situation familiar in the 3+1
spacetime. In this case, the function $\beta(r)$ is
simply not present in eq. (\ref{alphbet}). The only
surviving Einstein equation is (\ref{G00}) which now
takes the form
\be
\left(e^{-2\beta}\right)'=-2\varkappa\mu(r)r.
\ee
This possibility of existence of static dust solutions
in 2+1 can be related to the coefficient $D-3$ in
the weak-field approximation for $R^{(0)}_{(0)}$ and
in the equation for the ``Newtonian potential'' in this
case, see (\ref{2.1.2}) and (\ref{2.1.8}) respectively: 
the Newtonian potential simply vanishes in 2+1 for
any dust distribution.

It is clear that the equations obtained above, are
much more general that we needed in the description
of dust solutions. Indeed, they well serve in finding
many other solutions ({\em e.g.}, the 2+1 analogue
of the G\"odel rotating world); these results will
be presented elsewhere.

\subsection{Null solutions}

Solutions for null sources correspond in 3+1 to
those which describe spacetimes filled with
coherent radiation (null fluid; this latter name
is also applicable to {\em stationary} null fields,
in particular such types of electromagnetic and
fluid fields). In 3+1 this includes pp-wave
solutions, as well as the Robinson--Trautman null
radiation and Einstein--Maxwell fields, known only
in the absence of the cosmological term
\cite{ESEFE}. Here we find that in 2+1 this
limitation is lifted.

It is easy to see that the gravitational field
with
\be   \label{pp}
ds^2=A(u)y^2du^2+2dudv-dy^2
\ee
corresponds to
\be
R_{0202)}=R_{00}=R_{00}-
\frac{1}{2}g_{00}R=A(u), ~ R=0,
\ee
all other independent components of these
tensors being equal to zero. This is exactly the
2+1 Einstein--Euler pp-wave solution, its 3+1
counterpart being a direct product of this
2+1 spacetime and a one-dimensional space. It
is worth mentioning that the spacetime
(\ref{pp}) is conformally flat in the sense of
the Cotton--Schouten--York tensor, while its 3+1
counterpart is, of course, of type N.

A hybrid of (\ref{pp}) and the Robinson--Trautman
solution in the tetrad form
\be   \label{ppLambda}
\theta^{(0)}=e^{2\sqrt{\Lambda}y}du-\frac{A(u)
y^2}{2}dv, ~ \theta^{(1)}=dv, ~ \theta^{(2)}=dy,
\ee
\be    \label{seminull}
g_{(\alpha)(\beta)}=\left(
\begin{array}{ccc}
0 & 1 & 0\\
1 & 0 & 0\\
0 & 0 & -1\\
\end{array}
\right)
\ee
yields the non-zero independent components of
the Riemann--Christoffel tensor
\be
R_{(0)(1)(0)(1)}=R_{(0)(2)(1)(2)}=-\Lambda, ~
R_{(0)(2)(0)(2)}= A(u)(2\sqrt{\Lambda}y+1)
e^{-2\sqrt{\Lambda}y},
\ee
those of the Ricci curvature,
\be
R_{(0)(0)}=A(u)(2\sqrt{\Lambda}y+1)e^{-2
\sqrt{\Lambda}y}, ~ R_{(0)(1)}=-2\Lambda, ~
R_{(2)(2)}=2\Lambda,
\ee
the scalar curvature being $R=-6\Lambda$, and
non-zero components of the Einstein conservative
tensor being
\be
G_{(0)(0)}=A(u)(2\sqrt{\Lambda}y+1)e^{-2
\sqrt{\Lambda}y}, ~ G_{(0)(1)}=\Lambda, ~
G_{(2)(2)}=-\Lambda.
\ee
This is the same type of wave as (\ref{pp}), but
with the cosmological constant $\Lambda$. It is
obvious that when $\Lambda \rightarrow 0$ the
metric defined by (\ref{ppLambda}) and all its
concomitants reduce to those of (\ref{pp}).

A modification of the Robinson--Trautman
Einstein--Maxwell (radiation) solution to 2+1 in
the coordinates $u$, $r$, $\phi$ (the first two
of them pertaining to the null part of the basis)
is
\be      \label{RTLambda}
\theta^{(0)}=\left(-F(u)+\frac{1}{2}\Lambda r^2
\right)du+dr, ~ \theta^{(1)}=du, ~ \theta^{(2)}
=rd\phi.
\ee
Its non-zero independent concomitants are: the
Riemann--Christoffel tensor,
\be
R_{(0)(1)(0)(1)}=R_{(0)(2)(1)(2)}=-\Lambda, ~
R_{(1)(2)(1)(2)}=\frac{F'(u)}{r},
\ee
the Ricci tensor,
\be
R_{(0)(1)}=-R_{(2)(2)}=-2\Lambda, ~
R_{(1)(1)}=\frac{F'(u)}{r},
\ee
$R=-6\Lambda$, and Einstein's conservative tensor
\be
G_{(0)(1)}=-G_{(2)(2)}=\Lambda, ~
G_{(1)(1)}=\frac{F'(u)}{r}.
\ee
This now is a cosmological solution with null matter.

Moreover, the 3+1 Brdi\v{c}ka solution (see
\cite{ESEFE}, p. 236) which describes a
conformally flat spacetime with a constant null
Maxwell field (crossed electric and magnetic
fields with equal intensities), has a following
2+1 counterpart:
\be
\theta^{(0)}=du, ~ \theta^{(1)}=dv, ~
\theta^{(2)}=A(u)dy
\ee
with the same tetrad metric (\ref{seminull}).
In this case
\be
R_{(0)(2)(0)(2)}=R_{(0)(0)}=G_{(0)(0)}=
\frac{A''(u)}{A(u)}, ~ ~ R=0
\ee
(all other independent components vanish),
thus for $A(u)\sim\cosh(\omega u)$ with a
constant $\omega$, the same meaning as in
3+1 (only with the reinterpretation of 2+1
sources as fluids, perhaps a mixture of them)
persists, but for a general $A(u)$
this metric describes a wider class of null
solutions, including wavelike ones (however
without the $\Lambda$ term).

\setcounter{equation}{0}
\section{New prospects of the systematic
field-theo\-retic approach (conclusions)}%
\label{Sec8}

It is now possible to outline a scenario
of the most elementary stage of the universe
evolution from the very first step when
there are only two spacetime dimensions
($D=1+1$), the model which I had proposed
many years ago \cite{Mits81} but which has
taken a definitive shape only within the
theory formulated in this talk.

In 1+1, there is no real distinction between
space and time, so that the only invariant frame
is that of the light cone, the future and the
past being purely conditional. It is most
plausible that, similarly, only intrinsically
relativistic and null objects (fields)
can exist there. These are two fields: 0-form
and 1-form ones. From Table 1 we see that
they are the linear Maxwell-type and ghost
Machian fields, respectively. The
intrinsically relativistic 1-form Machian
field in 1+1 is arbitrary since it is
automatically a ghost field (the stress-energy
tensor identically vanishing, $\Lambda=0$).
Its intensity 2-form is proportional to the
2-dimensional Levi-Civit\`a symbol giving the
simplectic (skew) metric tensor, like that
which is used in the spinor (complex) 2-space.
It is still not clear if the proportionality
coefficient (an arbitrary function) pertains
to this metric or represents an independent
pseudoscalar object (the first possibility seems
to be more natural). These steps were realized
without any other introduction of a metric,
and now we get one directly from the theory,
nothing less than from the Machian field.
In the Maxwell-type field we have to use
exactly this metric. In 1+1, this is a linear
scalar field about which I have still nothing
more to tell. But to the metric {\em alias}
Machian field, while it is still completely
unconstrained, an additional action principle
could be applied, with a Lagrangian built of
the very metric and its derivatives.

We see that some, probably, very simple theory
works in 1+1. There should exist some mechanism
responsible for glueing together pairs of
elementary cells (areas) of 1+1 resulting in
elements of, most probably, the twistor space.
This glueing is not purely geometric, but more
a topological issue which results in a
four-dimensional geometry, obviously consistent
with our 3+1 spacetime.

\section*{Acknowledgements}

First of all, I wish to express my sincere thanks
to the Colleagues in the Organizing Committee of
the IV Workshop for dedicating it to my
seventy-years jubilee. I am greatly moved by
their friendly generosity so much shared by all
participants of the Workshop. My hope is that
God would permit me to live up to a fulfillment
of my responsibilities with respect to my
pupils, friends, and this hospitable country.

I am grateful to Alberto Garc{\'\i}a and Eloy
Ay\'on-Beato for fruitful discussions and
appreciate the hospitality of the CINVESTAV del
IPN (M\'exico D.F.) during my scientific visits.
My thanks are due to the participants of the IV
Workshop on Gravitation and Mathematical Physics
in Chapala, Jal., M\'exico, in particular to
D.V.~Ahluwalia, for their kind interest to this
talk given on November 26, 2001, and valuable
remarks.


\section*{Appendices}

\setcounter{section}{0}
\renewcommand{\thesection}{\Alph{section}}
\section{The weak field limit of Einstein's
equations with relativistic sources}\label{A}
\renewcommand{\theequation}{A.\arabic{equation}}
\setcounter{equation}{0}

The Newtonian approximation for the gravitational
field equation does not necessarily involve admission
of non-relativistic properties of the source terms
in Einstein's equations: it is sufficient to merely
consider the weak-field condition for gravitational
field. When a source has electromagnetic nature, one
simply {\em cannot} ignore its intrinsically
relativistic properties, since there cannot be
invented any non-relativistic approximation which
would describe electromagnetic stress-energy-momentum
complex adequately. 

Starting with Einstein's equations,
\be        \label{2.1.1a}
R^{(\alpha)}_{(\beta)}-\mbox{$\frac{1}{2}$}R
\delta^\alpha_\beta=-\varkappa
T^{(\alpha)}_{(\beta)},
\ee
and taking into account that
$R=\frac{2\varkappa}{D-2}T$,
we rewrite them in $D$ dimensions as
\be        \label{2.1.1b}
R^{(\alpha)}_{(\beta)}=-\varkappa\left(
T^{(\alpha)}_{(\beta)}
-\mbox{$\frac{1}{D-2}$}
T\delta^\alpha_\beta\right).
\ee
Thus for $00$-component we have
\be  \label{2.1.2}
R^{(0)}_{(0)}=-\mbox{$\frac{\varkappa}{D-2}$}
\left[(D-3)T^{(0)}_{(0)}-T^{(i)}_{(i)}\right].
\ee

For $D=4$ the temporal part of the
stress-energy tensor enters this equation
symmetrically with its spatial trace. Then
\be       \label{2.1.3}
R^{(0)}_{(0)}=-\mbox{$\frac{\varkappa}{2}$}
\left[T^{(0)}_{(0)}-T^{(i)}_{(i)}\right],
\ee
and for intrinsically relativistic sources
[$T^{(\alpha)}_{(\alpha)}=0$, the ($D-1$)-spatial
part $T^{(i)}_{(i)}$ has the same magnitude as
the temporal term $T^{(0)}_{(0)}$, but it comes
with the opposite sign]
\be  \label{2.1.4}
R^{(0)}_{(0)}=-\varkappa{T_{\rm 
intr.rel}}^{(0)}_{(0)}.
\ee
Taking into account that $\varkappa=8\pi G$,
where $G$ is the Newtonian gravitational constant
and the velocity of light $c=1$, we find that in
the intrinsically relativistic case $R^{(0)}_{(0)}$
is twice greater than in the non-relativistic case
(when $T^{(i)}_{(i)}\approx 0$).

For $D=3$ the temporal part is simply absent. But
when $T^{(\alpha)}_{(\alpha)}=0$, we have
$T^{(i)}_{(i)}=-T^{(0)}_{(0)}$ (the intrinsically
relativistic case). Thus there appears the same
coefficient $\varkappa=8\pi\gamma$, as it was the
case for $D=4$. This means that in 2+1 the
Newtonian-type potential can appear only in a
distribution of intrinsically relativistic matter
[see (\ref{2.1.2})]. Moreover, one has to keep in
mind the fact that there is no interaction between
islets of matter submerged in vacuum, so that in
order to come to an analogue of the Newtonian
potential one has to consider as the zeroth
approximation, an anti-de Sitter substratum.
However, an overall distribution of matter is also
admissible, especially when it asymptotically tends
to zero (to a vacuum, thus to the flat 2+1 spacetime),
and this seems to be more acceptable than the
cosmological term which inevitably does not lead
to such an asymptotic behaviour.

Let us consider here a static space-time with 
$g_{00}=1+2\Phi_{\rm N}$, $\Phi_{\rm N}\ll 1$.
The Newtonian approximation is now found from the
geodesic equation for a non-relativistic test
particle. One has to express $R^{(0)}_{(0)}$ in
terms of the Newtonian potential $\Phi_{\rm N}$
(in fact, its derivatives) neglecting the higher
order terms (non-linear in the Newtonian potential
and other corrections to the flat --- here,
Cartesian --- part of the metric coefficients in a
coordinated basis, all these corrections including
$\Phi_{\rm N}$ being considered as infinitesimals of
the first order of magnitude).

We choose now a static 1-form basis in spacetime,
\be   \label{2.1.5}
\theta^{(0)}={\rm e}^\alpha dt, ~ ~
\theta^{(k)}={g^{(k)}}_jdx^j,
\ee
this choice being here general enough. Then,
taking the inverse triad as ${g_{(k)}}^j$, so that
$dx^j={g_{(k)}}^j\theta^{(k)}$, $dt={\rm e}^{-\alpha}
\theta^{(0)}$, we find $ d\theta^{(0)}=\alpha_{,j}
{g_{(k)}}^j\theta^{(k)}\wedge\theta^{(0)}$, from
where it is easy to calculate the necessary
components of 1-form connections (in this static
case):
${\omega^{(0)}}_{(l)}\equiv{\omega^{(l)}}_{(0)}=
\alpha_{,j}{g_{(l)}}^j\theta^{(0)}$.
From the second Cartan structure equations,
$$
{\Omega^{(\alpha)}}_{(\beta)}=d{\omega^{(\alpha)}
}_{(\beta)}+{\omega^{(\alpha)}}_{(\gamma)}\wedge
{\omega^{(\gamma)}}_{(\beta)},
$$
neglecting non-linear terms (since in this
Appendix the weak-field approximation is
considered only), we find that
\be    \label{2.1.6}
{\Omega^{(0)}}_{(l)}\approx\left(\alpha_{,j}
{g_{(l)}}^j\right)_{,k}{g_{(i)}}^k\theta^{(i)}
\wedge\theta^{(0)}+\alpha_{,j}{g_{(l)}}^j
\alpha_{,i}{g_{(k)}}^i\theta^{(k)}\wedge\theta^{(0)}
\ee
(the last term is written here for symmetry reasons,
though it should be, of course, omitted). Now,
$$
R^{(0)}_{(0)}=g^{(l)(k)}{R^{(0)}}_{(l)(k)(0)}\approx
{\rm e}^{-\alpha}\left({\rm e}^\alpha\right)_{,i,j}
g^{ij}
$$
where $g^{ij}=-\delta^i_j ~ +$ higher-order terms
(to be neglected). Since ${\rm e}^\alpha\approx
1+\Phi_{\rm N}$,
\be    \label{2.1.7}
R^{(0)}_{(0)}\approx-\Delta\Phi_{\rm N}\approx
-\mbox{$\frac{\varkappa}{D-2}$}\left((D-3)T^{(0)}_{(0)}-
T^{(i)}_{(i)}\right), ~ ~ \varkappa=8\pi G;
\ee
thus, in the linear static approximation,
\be        \label{2.1.8}
\Delta\Phi_{\rm N}\approx\frac{8\pi G}{D-2}
\left((D-3)T^{(0)}_{(0)}-T^{(i)}_{(i)}\right),
\ee
the result coinciding in $D=4$ with (\ref{Poisson})
(the fields for which $-T^{(i)}_{(i)}\ll
T^{(0)}_{(0)}$, being non-relativistic, and for
which $-T^{(i)}_{(i)}/(D-1)\approx T^{(0)}_{(0)}$,
intrinsically relativistic). We see that for
arbitrary D in the rest frame of the fluid
the equation (\ref{2.1.8}) takes the form
\be        \label{2.1.8a}
\Delta\Phi_{\rm N}\approx\frac{8\pi G}{D-2}
\left[(D-3)\mu+(D-1)p\right].
\ee

When a perfect fluid is considered, its
energy-momentum tensor being (\ref{TmunuFl}), in
the rest frame of the fluid one has to compare
$(D-1)p$ and $(D-3)\mu$. The Newton--Poisson
equation (\ref{2.1.8}) takes for 3+1 the form
\be  \label{2.1.9}
\Delta\Phi_{\rm N}\approx 4\pi G(\mu+3p).
\ee
If $p\ll\mu$, the old traditional equation follows,
but if the fluid represents an incoherent radiation
($p=\mu/3$), the source doubles, and in the case of a
stiff matter ($p=\mu$), it quadruples. This last case
is, perhaps, not quite a physical one, as, probably,
all cases with $p>\mu/3$ which one may call
``hyperrelativistic'' ones. But the stiff matter case
may attract some attention since this is a simple model
which sometimes permits analytical consideration.

In 2+1 the situation changes drastically [see the
discussion of the expression (\ref{2.1.2}) above]:
the case when a usual Newtonian potential exists,
is shifted to intrinsically relativistic sources.
One has also to keep in mind that then the behaviour
of $\Phi_{{\rm N}}$ should correspond to a solution
of the Poisson equation with the two-dimensional
Laplacian. This is, of course, not applicable to the
``stiff matter'' in 2+1 (when $p=\mu$), since the only
exact solution of Einstein's equations existing in this
case, is that with a cosmological term which is
constant, thus excluding the flat spacetime asymptotics.

The weak field approximation, of course, does not
affect exact results in general relativity, in
particular, in cosmology. However, some traditional
principles of physics are sensitive to the approximate
forms of equations, and one of the most important
examples is the principle of equivalence. The
conclusions drawn in this Appendix suggest a
relativistic generalization of this principle,
especially since the Newtonian-type potential is
generated by a wide class of distribution of matter,
including intrinsically relativistic and
hyperrelativistic matter: the only restriction in
this case consists of the weakness of the field and
not the ``state of motion'' of the sources in
Einstein's equations (especially such an intrinsic
property as to be relativistic which is so often
realized by static configurations when the very idea
of motion is out of the question). As to the
applications of this generalized principle of
equivalence, it is worth pointing out the (post-)
post-Newtonian approximations. Since some conclusions
about validity of the principle of equivalence come
from observations of stellar systems, a mere presence
in them of intrinsically relativistic objects (say,
high density of any kind of radiation, strong or
widely distributed magnetic fields, existence of
stiff matter in cores of exotic stars) would radically
change interpretation of the observational data if
their proper understanding depends on adequate
application of approximated description.

\section{Intrinsically relativistic fields}\label{B}
\renewcommand{\theequation}{B.\arabic{equation}}
\setcounter{equation}{0}

We introduce here in the general case of $D=n+1$
spacetime the concept of intrinsically relativistic
objects which remain relativistic even when being
``at rest'' (static or stationary, when fields and
not particles are considered). This concept was
already used implicitly in the four-dimensional case,
especially when a perfect fluid with the equation of
state $p=\mu/3$ (incoherent radiation) was considered.
One of the reasons to take seriously the concept of
intrinsically relativistic fields (and objects)
consists of appearance of factor 2 in effects
of their interaction with weak gravitational fields
in 3+1 dimensions. This factor was first noticed in
a comparative study of the effect of bending of light
rays in the gravitational field (of sun): Soldner
\cite{Soldn} and Einstein \cite{Einst1} versus Einstein
\cite{Einst2} (see \cite{Tolman, Somm52, MisThorWhe,
Weinb72}). There exists, of course, also the ``inverse''
(in the spirit of the Newtonian third law) effect
(generation of gravitational field by electromagnetic
field) involving doubled electromagnetic energy density
(\cite{Mitsdep, VlaMitsHor, Mits91, Mits92}) in the
four-dimensional weak gravitational field approximation
(see Appendix A):
\be \label{Poisson}
\Delta\Phi_{\rm N}=4\pi G\mu_{\rm non-relat}+
8\pi G\mu_{\rm em}
\ee
[this (at least) doubling occurs, of course, for any
intrinsically relativistic sources, not only Maxwell's
field in 3+1]. Thus the intrinsically relativistic
objects are exceptional only in the sense that their
relativistic nature is absolute and does not depend on
the choice of reference frame in which they are observed;
a concrete consideration of these properties for
arbitrary dimensionality of spacetime see below.

If a single point-like object is intrinsically
relativistic, it has to move with the speed of light
(a null world line, ${\mbox{\bf p}}^2=E^2$), since only
this velocity is absolute (both in the special and
general relativity). For a distributed matter (in
particular, a field), this corresponds to vanishing of
the trace of its stress-energy tensor: in certain sense,
temporal and spatial parts of its energy-momentum
tensor contribute equally, but with opposite signs.

The stress-energy tensor of an $r$-form field
in general takes the form (\ref{2.1}), since
the Lagrangian density, as well as the function
$L$ depend on $g^{\mu\nu}$ only algebraically
(the $r$-form potentials are considered to be
independent of the metric tensor). The trace
of this stress-energy tensor is (\ref{2.2}).
Then the intrinsically relativistic property
condition $T^\alpha_\alpha=0$ yields $L\sim I^k$,
$k=\frac{D}{2(r+1)}$. Another way to deduce this
expression for $k$, if the homogeneity law
$L\sim I^k$ is already accepted, consists of
equally distributing the metric tensor factors
(including those which are found in $\sqrt{|g|}$)
in the definition of ${\frak L}=\sqrt{|g|}L$
between the field tensor components of the
$r$-form field, with the subsequent application
of the Noether theorem \cite{Mits58, Mits69}:
\be  \label{1.1.2}
{\frak T}^\beta_\alpha:=\frac{\delta{\frak
 L}}{\delta g_{\mu\nu}}g_{\mu\nu}|^\beta_\alpha
\equiv \frac{\delta{\frak L}}{\delta g^{\mu\nu}}
g^{\mu\nu}|^\beta_\alpha=\frac{\delta{\frak L}
}{\delta\left(|g|^{\frac{1}{2(r+1)}}g^{\mu\nu}
\right)}\left.\left(|g|^{\frac{1}{2(r+1)}}
g^{\mu\nu}\right)\right|^\beta_\alpha.
\ee
Then it is clear that the intrinsically relativistic
property of the field is equivalent to vanishing of
trace of the Trautman coefficient \cite{Traut}
$\left.\left(|g|^{\frac{1}{2(r+1)}}g^{\mu\nu}
\right)\right|^\beta_\alpha$:
\be   \label{2.2.2}
\left.\left(|g|^{\frac{1}{2(r+1)}}g^{\mu\nu}\right)
\right|^\alpha_\alpha=|g|^{\frac{1}{2(r+1)}}
g^{\mu\nu}\left(2-\frac{D}{k(r+1)}\right)=
\mbox{ (the ansatz) }=0,
\ee
{\em quod erat demonstrandum}.
When $k=1$, only space-times of even number of
dimensions $D$ can fit this condition: $D=2(r+1)$.
The same condition determines the conformal
invariance property of the fields.

Thus in the intrinsically relativistic case it is
necessary and sufficient to use the simplest
nonlinear Lagrangian densities (see the Table 1),
\be  \label{2.3.1}
{\frak L}=\sqrt{|g|}\sigma I^k,  ~ ~ ~
k=\frac{D}{2(r+1)}.
\ee

\begin{center}
\begin{tabular}{|l||c|c|c|c|c|c|c|} \hline 
$r\backslash D$ & 2 & 3 & 4 & 5 & 6 & 7 & 8 \\
 \hline \hline
0 & 1 & 3/2 & 2 & 5/2 & 3 & 7/2 & 4 \\ \hline
1 & 1/2 & 3/4 & 1 & 5/4 & 3/2 & 7/4 & 2 \\ \hline
2 & ~ & 1/2 & 2/3 & 5/6 & 1 & 7/6 & 4/3 \\ \hline
3 & ~ & ~ & 1/2 & 5/8 & 3/4 & 7/8 & 1 \\ \hline
4 & ~ & ~ & ~ &   1/2 & 3/5 & 7/10 & 4/5 \\ \hline
5 & ~ & ~ & ~ &   ~ &   1/2 & 7/12 & 2/3 \\ \hline
6 & ~ & ~ & ~ & ~ &     ~   & 1/2 & 4/7 \\ \hline
7 & ~ & ~ & ~ & ~ & ~ & ~ & 1/2 \\ \hline
\end{tabular} \\
~~~~~~ \\
Table 1. Values of $k$ versus $r$ and $D$,
 describing\\ general intrinsically relativistic
 fields (\ref{2.3.1}).
\end{center}

This table simply gives values of $k$; since
$0\leqslant r\leqslant D-1$, the lower left corner
consists of blank spaces only: the ``missing''
$r$-form fields are either trivially exact ones,
or equal to zero.

Now one may consider intrinsically relativistic
fields of any rank $r$ for every dimension $D$,
this being possible at the cost of admission of
non-linear fields ($k\neq1$). When $k=1$, a linear
field is realized ({\em cf.} the Table 1). It is
easy to see that all these intrinsically
relativistic fields (for all corresponding values
of $D$) automatically possess the property of
conformal invariance. When $T^\alpha_\alpha<0$
(fields which are ``more relativistic'' than,
for example, the 3+1 Maxwell field and the
incoherent radiation are), we can speak on
intrinsically hyperrelativistic fields,
corresponding in their ultimate case to the stiff
matter.

The concept of a fundamental field in $D$
dimensions analogous to the 3+1 Maxwell field,
can be now formulated as that of a linear
intrinsically relativistic field. Thus in
odd-dimensional spacetimes there is no room for
Maxwell-like fields (for example, in 2+1 there
is no analogue of the electromagnetic field in
its proper sense), and in the even-dimensional
ones, such fields generally should not be
described by a 1-form potential (which is the case
in 3+1 only). In 5+1-dimensional spacetime, this
will be a 2-form field; in 7+1, 3-form field;
in 9+1, 4-form field, and so on. They all admit
the dual conjugation of the respective field
tensor (since its rank is $D/2$), yielding
relations familiar from Maxwell's theory,
though they involve objects of other ranks.
These two distinctive properties (linearity and
intrinsically relativistic one) seem to be of a
great physical importance, which single out these
fields from many others (but probably not from the
Machian $r=D-1$ fields filling the lower nontrivial
diagonal in the Table 1). The Machian intrinsically
relativistic fields correspond to $k=1/2$, hence to
$\Lambda=0$; their components are in fact arbitrary
functions of spacetime coordinates, thus such fields
differ quite radically from all other physical fields.

In particular, these results yield a\\
{\bf Theorem}: (Generalized) electromagnetic fields
exist only in even $D$ spacetime dimensions, then
being $(r=D/2-1)$-form fields. They possess all
essential properties of the 3+1 Maxwell fields
(are linear, intrinsically relativistic, and subject
to the $D$-dimensional dual conjugation relations).

All other fields in the Table 1 seem to be of less
general importance; for example, the ($r=D-2$)-field
models perfect fluids in the respective spacetime,
and its Lagrangian needs to be chosen as such a
function of the field invariant which yields the
desired equation of state.

\end{document}